\documentclass[prb,a4,twocolumn,aps,superscriptaddress,showpacs,floatfix]{revtex4}

\usepackage{graphicx}
\usepackage{amsmath}
\usepackage{amsfonts}
\usepackage{paralist}
\begin{document}

\def\be{\begin{equation}}
\def\ba{\begin{eqnarray}}
\def\ee#1{\label{#1}\end{equation}}
\def\ea#1{\label{#1}\end{eqnarray}}
\def\la{\langle \! \langle}
\def\ra{\rangle \! \rangle}
\def\bs{\begin{center}}
\def\es{\end{center}}
\def\fpa#1{\frac{\partial}{\partial #1}}
\def\ve{\varepsilon}
\def\De{D_{eff}}
\def\td{\textrm{d}}
\def\he{\hbar/2e}
\graphicspath{{img/}}
\title{Anomalous transport in biased ac-driven Josephson junctions: Negative conductances}

\author{M. Kostur}
\affiliation{Institute of Physics, University of Silesia,
40-007 Katowice, Poland}
\author{L. Machura}
\affiliation{Institute of Physics, University of Silesia,
40-007 Katowice, Poland}
\author{P. Talkner}
\affiliation{Institute of Physics, University of Augsburg,
D-86135 Augsburg, Germany}
\author{P. H\"anggi}
\affiliation{Institute of Physics, University of Augsburg,
D-86135 Augsburg, Germany}
\author{J. \L uczka}
\affiliation{Institute of Physics,  University of Silesia,
40-007 Katowice, Poland}

\begin{abstract}

We investigate classical anomalous electrical transport in a driven,
resistively and capacitively shunted Josephson junction device.
Novel transport phenomena are identified in chaotic regimes when the
junction is subjected to both, a time periodic (ac) and a constant,
biasing (dc) current. The dependence of the voltage across the
junction on the dc-current exhibits  a rich diversity of anomalous
transport characteristics: In particular, depending on the chosen
parameter regime we can identify  so termed absolute negative
conductance around zero dc-bias, the occurrence of negative
differential conductance and, after crossing a zero conductance, the
emergence of a negative nonlinear conductance in the non-equilibrium
response regime remote from zero dc-bias.

\end{abstract}

\maketitle

\section{Introduction}

The  Josephson junction system constitutes a beautiful paradigm of a
nonlinear system exhibiting most interesting classical and quantum
phenomena.  \cite{junction} This system in addition offers a rich
spectrum of beneficial  applications. For example, a prominent
application of a Josephson junction set-up relates to the definition
of the voltage standard. Moreover, practical devices based upon the
characteristics of a Josephson junction are instrumental for high
speed circuits: They can be designed to switch voltage within  a few
picoseconds. Their typical attribute of a low power dissipation
proves them serviceable in high-density computer circuits where the
resistive heating limits the applicability of conventional switches.
More recent applications refer to quantum computing devices where
Josephson junction set-ups can store single units of information
(qubits), \cite{qubits} or with appropriately engineered coupling
among various units, they serve as an architecture for the
processing of quantum information. \cite{capa}  Yet,  there remain
still new phenomena to be uncovered for this system which in turn
carry the potential for new applications.  Of special interest are
novel transport phenomena in presence of external ac- and
dc-forcing. In this context, the phenomenon of  absolute negative
conductance (ANC) plays a particularly intriguing role. Here, ANC
means that the system's response is {\it opposite} to a {\it small}
external bias. In the present context, a  small, say, positive
dc-current generates  a negative voltage across the junction. This
constitutes no contradiction to thermodynamic laws because it occurs
in presence of simultaneous acting non-equilibrium perturbation, the
ac-drive. This phenomenon is known to emerge within a quantum
mechanical setting in presence of tunneling processes; remarkably,
however, it also has been demonstrated on a classical level in
stylized, spatially extended ratchet-like systems without reflection
symmetry. \cite{BroBen2000,EicRei2002a,EicRei2002b,CleBro2002}

The phenomenon of ANC has already been experimentally observed in
p-modulation-doped GaAs quantum wells \cite{hop} and also in
semiconductor superlattices, occurring therein as a genuine quantum
phenomenon. \cite{keay} Interestingly enough, this very
ANC-phenomenon has been reported  first in a recent work by us in
Ref. \onlinecite{machura} for an ac- and dc-driven Josephson
junction. Notably, however, the study in Ref.  \onlinecite{machura}
identifies this phenomenon within its {\it classical} operation
regime for an inherent  reflection-symmetric  system. In extending
our previous study in Ref. \onlinecite{machura}, we (i) provide further
details, (ii) explore even wider regimes in parameter space and
(iii) identify additional novel response regimes. In contrast to a
related study, \cite{speer} where ANC is investigated in an
underdamped, deterministic chaos regime, here we emphasize the role
of thermal noise and noise-induced nonlinear response phenomena.
Moreover, we  put forward a study of an accessible, optimal
parameter regime towards the objective for an experimental
verification of our findings. The underlying dynamics of this driven
Josephson junction can conveniently be described by the model of a
resistively and capacitively shunted junction in terms of the so
called Stewart-McCumber model. \cite{junction,kautz}

The layout of the present work is  as follows: In Sec. II, we
present the classical Stewart-McCumber model of the Josephson
junction. Next, in Sec. III, we elucidate various  regimes of
anomalous transport behavior, such as ANC,  negative differential
conductance (NDC) and  so termed negative-valued nonlinear
conductance (NNC).

In Sec. IV, we work out the optimal regimes for the phenomenon of
the negative conductance. In Sec. V, we elaborate on the
experimental feasibility for an in situ confirmation of our
diversified novel theoretical predictions. Sec. VI  provides a
summary and some conclusions.


\section{The Stewart-McCumber model}

This model describes  the (semi)-classical regime of the
voltage-current characteristics of a Josephson junction. The model
involves the Josephson supercurrent characterized by the critical
current $I_0$, a normal  (Ohmic) current characterized by the normal
state resistance $R$ and a displacement current accompanied with the
junction capacitance $C$. The ubiquitous thermal equilibrium noise
consists of  Johnson noise associated with the resistance $R$. The
dynamics of the phase difference $\phi=\phi(t)$ across the junction
is then described by the following nonlinear equation of motion,
e.g. see in the review \onlinecite{kautz},
\begin{eqnarray} \label{JJ1}
\Big( \frac{\hbar}{2e} \Big)^2 C\:\ddot{\phi} + \Big( \frac{\hbar}{2e} \Big)^2 \frac{1}{R} \dot{\phi}
+ \frac{\hbar}{2e} I_0 \sin (\phi) = \frac{\hbar}{2e} I_d \nonumber\\
+ \frac{\hbar}{2e}I_a \cos(\Omega t) + \frac{\hbar}{2e}
\sqrt{\frac{2 k_B T}{R}} \:\xi (t)\;.
\end{eqnarray}
Herein, a dot denotes the differentiation with respect to time $t$,
$I_d$ and $I_a$ are the amplitudes of the applied dc- and
ac-currents, respectively,
 $\Omega$ is the angular frequency of the ac-driving.  The parameter $k_B$
denotes the Boltzmann constant and $T$ is the temperature of the
system. Thermal equilibrium fluctuations are modeled by
$\delta$-correlated Gaussian white noise $\xi(t)$ of zero mean and
unit intensity, i.e.,  $<\xi(t) \xi(s)> = \delta(t-s)$.

The limitations of the Stewart-McCumber model and its range of
validity are discussed e.g. in Sec. 2.5 and 2.6  of  the
comprehensive review  paper by Kautz. \cite{kautz} In particular, we
thus work within the small junction area limit and in a regime where
photon-assisted tunneling phenomena do not contribute. Throughout
the following we shall adopt the dimensionless form of Eq.
(\ref{JJ1}) from Refs. \onlinecite{junction} and \onlinecite{kautz},
namely,
\begin{eqnarray}
\label{JJ2} \frac{d^2 \phi}{dt'^2} + {\sigma} \frac {d  \phi}{d t'}
+ \sin (\phi)  = i_0 + i_1  \cos(\Omega_1 t')
+ \sqrt{2 \sigma D} \; \Gamma(t'), \nonumber \\
\end{eqnarray}
where the 
dimensionless time $t'=t/\tau_0$ and  the characteristic time $\tau_0 = 1/\omega_p$. The
 Josephson plasma frequency $\omega_p=(1/\hbar)\sqrt{8 E_JE_C}$
is expressed by the Josephson coupling energy $E_J=(\hbar/2e)I_0$
and the charging energy  $E_C=e^2/2 C$.
The 'friction' coefficient  ${\sigma} = \tau_0 / RC$ is given by the
ratio of two characteristic times: $\tau_0$ and the relaxation time
$\tau_r = RC$. This dimensionless friction parameter $\sigma$ measures the
strength of dissipation. The amplitude and the angular frequency of
the ac-current are $i_1 =  I_a / I_0$ and $\Omega_1 = \Omega
\tau_0=\Omega/\omega_p$, respectively.  The re-scaled dc-current
reads $i_0=  I_d/ I_0$, the re-scaled zero-mean Gaussian white noise
$\Gamma(t')$ possesses the auto-correlation function $\langle
\Gamma(t')\Gamma(u)\rangle=\delta(t'-u)$, and the noise intensity $D
= k_B T / E_J$ is given as the ratio of two energies, the thermal
energy and the Josephson coupling energy (corresponding to the
barrier height). A different scaling procedure, being more familiar
within the Brownian motor community, is detailed in Appendix.

The most important characteristic for the above system is the
current-voltage curve. To obtain it, we numerically integrated Eq.
(\ref{JJ2}). For long time (to avoid initial conditions and
transient effects), we calculated the stationary  dimensionless
voltage
\begin{eqnarray}
\label{v}
v=\langle \frac{d \phi}{dt'} \rangle,
\end{eqnarray}
where the brackets  denote an average over both,  all the
realizations of the thermal noise and a temporal average over one
cycle period of the external ac-driving. The stationary physical
voltage is then expressed as
\begin{equation}
\label{V}
V= \frac{\hbar \omega_p}{2e} \; v.
\end{equation}
Strictly speaking, the zero temperature  limit, $D=0$, should not be
considered within the framework of the  Stewart-McCumber model;
because at zero temperature the quantum dynamics comes into play.
However, in order to explain peculiar properties of the system at
non-zero temperature and/or to seek
 optimal regimes, it is insightful nevertheless to consider the
deterministic dynamics, i.e. the  zero noise case,  $D=0$. Then, the
role of initial conditions are also relevant. This is particularly
so if several attractors coexist. In such a case,  an additional
average over initial conditions must be   performed: We have chosen
initial phases $\phi_0$ that are equally distributed over one period
$[0, 2\pi]$ and (dimensionless) initial voltages  $v_0$
 equally distributed in the range $[-2, 2]$. Further details of how
to treat the deterministic case are presented e.g. in Ref.
\onlinecite{family}.  In the case of several attractors, it means
that the average is  over attractors,  whose weights are
proportional to corresponding basins of attractions.   However, it
is worth to stress that the case $D=0$ generally  is not
equivalent to  the limit $D \to 0$  performed in the non-zero
temperature, Fokker-Planck case, see the studies in Ref.
\onlinecite{Junghanggi}.

Physical systems described by Eq. (\ref{JJ2}) are widespread and
well known. An example is  a Brownian particle moving in the
spatially periodic potential $U(\phi)=U(\phi+L)= - \cos( \phi)$ of
period $L= 2\pi$, driven by the time-periodic force and a constant
force.  \cite{MachuraJPC}  Then, the variable $\phi$ corresponds to
the space coordinate of the Brownian particle and the ac- and
dc-currents plays the role of driving forces on the particle. Other
specific systems include rotating dipoles in external fields,
\cite{Reg2000, Coffey} superionic conductors \cite{Ful1975} or
charge density waves, \cite{Gru1981}  to name just a few.

\section{ Anomalous transport behavior}

The deterministic dynamics corresponding to (\ref{JJ2}) encompasses a
three-dimensional phase space, namely $\{\phi, d \phi/dt',  \Omega_1 t'\}$
and contains four parameters, reading $\{\sigma, i_0, i_1,
\Omega_1\}$. Therefore, its dynamics is able to exhibit  an
extremely rich behavior in phase space as a function of the chosen
system parameters. \cite{kautz,salerno} For example, its dynamics
features harmonic, subharmonic, quasiperiodic and also chaotic types
of behavior;  for further details we refer the readers to the review in Ref.
\onlinecite{kautz}. At non-zero temperature, $D>0$, thermal
fluctuations lead to diffusive dynamics for which stochastic escape
events among possibly coexisting attractors are possible.
\cite{RevMod}
Moreover, when thermal noise is acting, the system dynamics is
constantly excited away from stable trajectories; it  thus can
explore the whole phase space. In some cases the probability
distribution can concentrate on regions in phase space which do not
coincide with stable trajectories. The prominent example of such
phenomena is a classical excitable stable point, which under
influence of noise can produce a ``limit cycle''-like probability
distributions.  \cite{KosSai2003}   This in turn  can imply drastic
consequences for the transport properties.

From the reflection symmetry $\phi \to -\phi$ of the potential
$U(\phi) = -\cos(\phi)$ and the time-reflection $t'\to -t'$ of the
ac-driving, it follows that the average voltage is strictly {\it
zero} if the dc-current is zero ($i_0=0$). If a nonvanishing
dc-current is applied, however, (i.e. $i_0\ne 0)$, the above
mentioned reflection symmetries are broken and a non-zero voltage
typically emerges. Since the dynamics determined by Eq. (\ref{JJ2})
is nonlinear and the system is multidimensional, it should not come
as a surprise that the current-voltage  characteristics is typically
nonlinear and often depicts a non-monotonic function of the system
parameters. Nevertheless, some most non-intuitive behaviors still
remain to be unraveled in parameter space which seemingly have
escaped previous detailed investigations for this archetype system
\cite{kautz}.
\begin{figure}[htb]
\includegraphics[angle=0,width=0.95\linewidth]{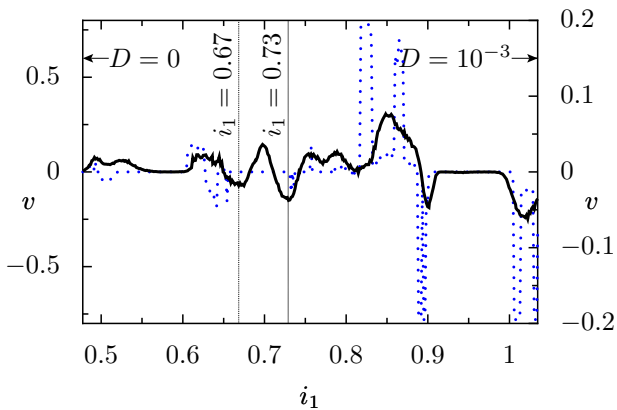}
\caption{ The diagram depicts the dependence of the voltage versus
the variation of the driving
  amplitude $i_1$ of the ac-current for a positive dc-current set at
  $i_0=0.0159$. Remaining  parameters are  chosen as $\sigma=0.143$ , $\Omega_1=0.78$ and
  temperature $D=0$ (dotted line, blue online, left scale) and $D=10^{-3}$ (solid
  line, black online, right scale).  }
\label{adm}
\end{figure}

The current-voltage curve $v=v(i_0) $  is  a nonlinear function
 of the dimensionless dc-current strength $i_0$.
Upon inspecting the symmetries in the equation of motion  we find
that  this function is odd in the dc-bias $i_0$,  i.e. $v(-i_0)
=-v(i_0)$. Typically, the voltage is an increasing function of the
dc-current.  
Such regimes correspond in parameter space  to a normal, Ohmic-like
transport behavior. More interesting are, however, those regimes of
anomalous transport, exhibiting (i) an (absolute) negative
conductance  near zero dc-bias, (ii) the regime of a negative
differential conductance, and (iii)  after crossing zero
conductance, a negative-valued conductance in a nonlinear regime
displaced from zero dc-bias $i_0=0$.

The numerical analysis of the system dynamics depicts that anomalous
transport occurs in a parameter range where the driven system
dynamics is strongly nonlinear. Although there is  no obvious direct
connection to  chaotic properties of the system dynamics we have
found that regimes of anomalous transport typically necessitate also
a chaotic dynamics.  In particular, for the regime $\sigma=0.143$
and $\Omega_1=0.78$ presented in Fig. \ref{adm}, the numerical study
shows that  the first bifurcation cascade at $D=0$ leading to chaos
occurs near $i_1=0.477 $, c.f. the dotted line in Fig. \ref{adm},
while the regime of an anomalous transport behavior starts out  near
$i_1 \simeq 0.637$. Not unexpectedly, no anomalous transport occurs
in the regime of approximate linear dynamics.
\begin{figure}[htb]
\includegraphics[angle=0,width=0.95\linewidth]{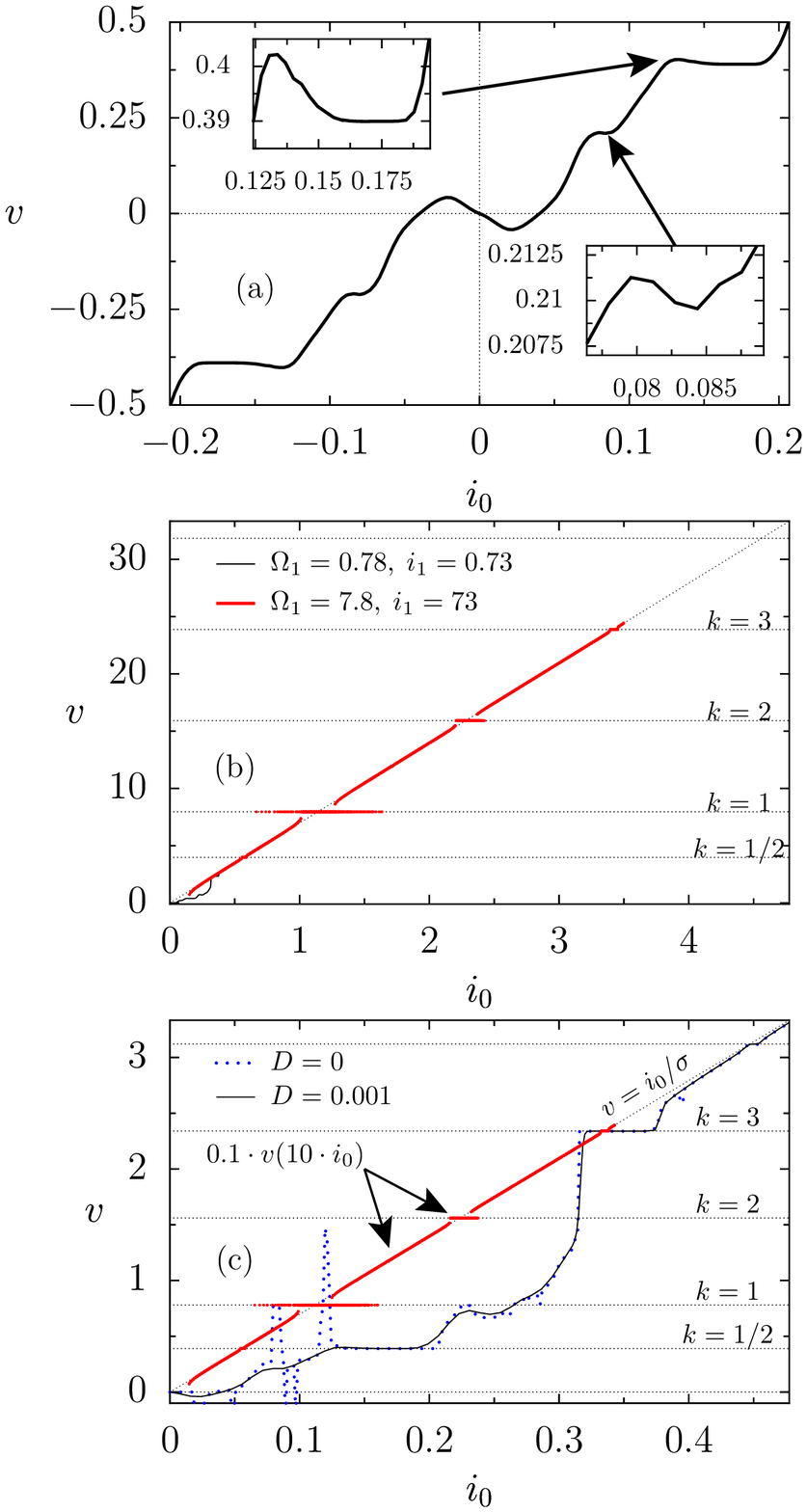}
\caption{The functional dependence of voltage $v$ on the dc current
  $i_{0}$ (i-v characteristic) is compared for two sets of
  parameters which either display negative conductance or Shapiro steps.
Panel (a) which corresponds to ac frequency $\Omega_{1} = 0.78$, ac driving
strength $i_{1}=0.73$, friction $\sigma=0.143$ and noise strength $D=10^{-3}$
displays both absolute negative mobility in an interval containing $i_{0}=0$
and different regions with negative differential conductivity,
cf. also the respective enlarged regions in the two insets. 
Panel (b) displays the i-v characteristic for the parameter values of
panel (a) (thin solid line) and the i-v characteristic 
for $\Omega_1=7.8$, $i_{1}=73$, $\sigma = 0.143$ and $D=0$ (thick solid
line). Shapiro steps are 
clearly visible at
$v=k i_{0}/\sigma$ for $k=1/2,1,2,3$ (dotted horizontal lines). 
Away from these steps, the i-v
charateristic follows the Ohmic law $v=i_{0}/\sigma$ (dotted inclined line).   
Panel (c) presents the i-v characteristic for the small driving
frequency ($\Omega_1=0.78$) for positive bias currents $i_{0}$ up to
$i_{0}=0.48$ (solid line), the i-v characteristic for the same set of
junction parameters at zero noise ($D=0$) (dotted line with wide
distance of dots) and compares them with a rescaled version  $v_{c}:= 0.1\cdot
v(10\cdot i_{0})$ of the i-v
characteristic for the larger frequency such that the full range shown
in panel (b) is displayed again. 
Whereas the linear overall behavior of the large frequency curve is
interrupted only at a few Shapiro steps, the small frequency i-v
characteristic is dominated by the nonlinearities also away from the
locking regions at $v=\Omega_{1}/2$ and $v=3 \Omega_{1}$.}
  \label{ndm}
\end{figure}
\subsection{Absolute negative  conductance}

In  transport theory, the Green-Kubo linear response regime plays an
important role.  It  allows one to obtain linear transport
coefficients. For the system described by Eq. (\ref{JJ2}), there are
regimes, where for sufficiently small values of the dc-bias  $i_0$,
linear response theory holds. This is characterized by the relation
\cite{risken,ustinov}
\begin{equation}
\label{LR}
v= r_L  i_0.
\end{equation}
It defines the linear transport  coefficient  $r_L=r_L(\sigma, i_1,
\Omega_1, D) $ which does not depend on $i_0$ and is    called  the
static  resistance  or $g_L=1/r_L$ is the  conductance in the linear
response regime. Since the full response exhibits also a non-linear
behavior, it is necessary to recall that the resistance $r_L$ is
defined in the limit of a  small dc-current, $r_L= \lim_{i_0 \to 0}
[v(i_0)/i_0]$.  The case $r_L > 0$ corresponds to the normal
transport behavior or the
 Ohmic-like  regime. The case  $r_L < 0$
 amounts to the situation when the voltage assumes the opposite sign of the dc-bias. It is
termed here {\it  absolute negative conductance} (in analogy to the
absolute negative mobility of driven Brownian particles
\cite{BroBen2000,EicRei2002a,EicRei2002b,CleBro2002}).
In Fig. \ref{ndm}  we exemplify this situation.  Indeed, for
dc-current values varying between $i_0 \in (-0.04, 0.04)$,  the
voltage assumes a sign opposite to $i_0$. The driving amplitude of
the ac-current at $i_1=0.73$ corresponds to the second
negative-valued minimum of the averaged voltage shown in Fig.
\ref{adm} for a temperature $D=10^{-3}$. The first  negative-valued
minimum of $v$ versus $i_1$ in Fig. \ref{adm} has been  explored in
greater detail  in our previous work  in Ref. \onlinecite{machura}.
Let us emphasize here that, in distinct difference to this case with
non-zero thermal noise $D\neq 0$, in the limit of vanishing noise
(i.e. when $D=0$) the averaged voltage is in fact identically zero
at the ac-driving strength $i_1=0.73$. This implies that this
counter-intuitive phenomenon of ANC is solely induced by  {\it
thermal equilibrium fluctuations}. For the record, there exist also
linear response regimes where even in the deterministic case, the
voltage can become negative for  a positive dc-current (and vice
verse). \cite{machura,speer}  From Fig. \ref{adm}, one can observe
that there exist several parameter windows depicting such a
noise-induced ANC-behavior: the conductance $g_L=g_L(\sigma, i_1,
\Omega_1, D) $ exhibits sign changes multiple times upon increasing
the ac-amplitude strength $i_1$. Between such ANC windows with $g_L
< 0$, normal  transport regimes with $g_L > 0 $ occur. Moreover, we
tested that the ANC windows remain stable upon a small variation  of
the remaining parameters.

\subsection{Negative differential conductance}

If the voltage $v$ is not a  monotonic function of the dc-current
$i_0$,  the {\it differential} conductance  can assume
negative values. The differential (or dynamic) resistance, defined
by the relation \cite{ustinov}
\begin{equation}
\label{DR}
 r_D(i_0) = \frac{d v(i_0)}{d i_0} ,
\end{equation}
 may therefore assume negative values within some interval of dc-bias values  $i_0$.
In Fig. \ref{ndm}, we depict two  examples of such behavior: the
negative differential conductance (NDC)  $g_D(i_0) =1/r_D(i_0)$ is detected  for the
dc-current $i_0 \in (0.0796,0.086)$ and $i_0 \in (0.132,0.166)$, note the two 
inset panels in Fig. \ref{ndm}. To the
best of our knowledge, this typical NDC voltage-current
characteristic shown in Fig. \ref{ndm} has not yet  been
experimentally reported for symmetrical systems, such as the system
in Eq. (\ref{JJ2}). For asymmetric ratchet systems, this effect was
described in Ref. \onlinecite{machuraA}.

We note that both effects of absolute and differential negative conductance 
are found in the part of the  $\Omega_{1}-\sigma$ plane where
$\Omega_{1}<1$ and $\Omega_{1} > \sigma$. In this region 
Shapiro steps do not exist, i.e. there are no phase locked regions 
which interrupt an otherwise linear current current-voltage
characteristic. 
For a detailed analysis of
the different regions of the $\Omega_{1}-\sigma$ plane with respect to
the occurence of Shapiro steps see the
discussion 
in chapter 5.1 of the review by Kautz\cite{kautz}. In Fig.~\ref{ndm}
current voltage characteristics for two parameters sets 
$\Omega_{1}= 0.78$, $\sigma =
0.143$ and $\Omega_{1}= 7.8$, 
$\sigma=0.143$ are displayed. 
The characteristics for the smaller frequency  contain several bias current
intervals with negative conductance, several locking regimes  
but no Shapiro steps. The second current voltage characteristic does have
pronounced Shapiro steps but no regions with negative conductance.

\subsection{ Nonlinear response regime displaced from zero 
bias $i_0=0$: negative-valued  conductance}

 In the nonlinear response regime, the "nonlinear resistance" or the static
 resistance at a fixed bias current is  defined by the relation \cite{risken,ustinov}
\begin{equation}
\label{NR} r_N (i_0) = r_N(i_0; \sigma, i_1, \Omega_1, D)  =
\frac{v(i_0)}{i_0} \;.
\end{equation}
It typically depends on $i_0$ in a nonlinear and non-monotonic
manner. The nonlinear conductance refers to its  inverse, i.e.,
$g_N(i_0)=1/r_N(i_0)$.
\begin{figure}[htb]
\includegraphics[angle=0,width=0.95\linewidth ]{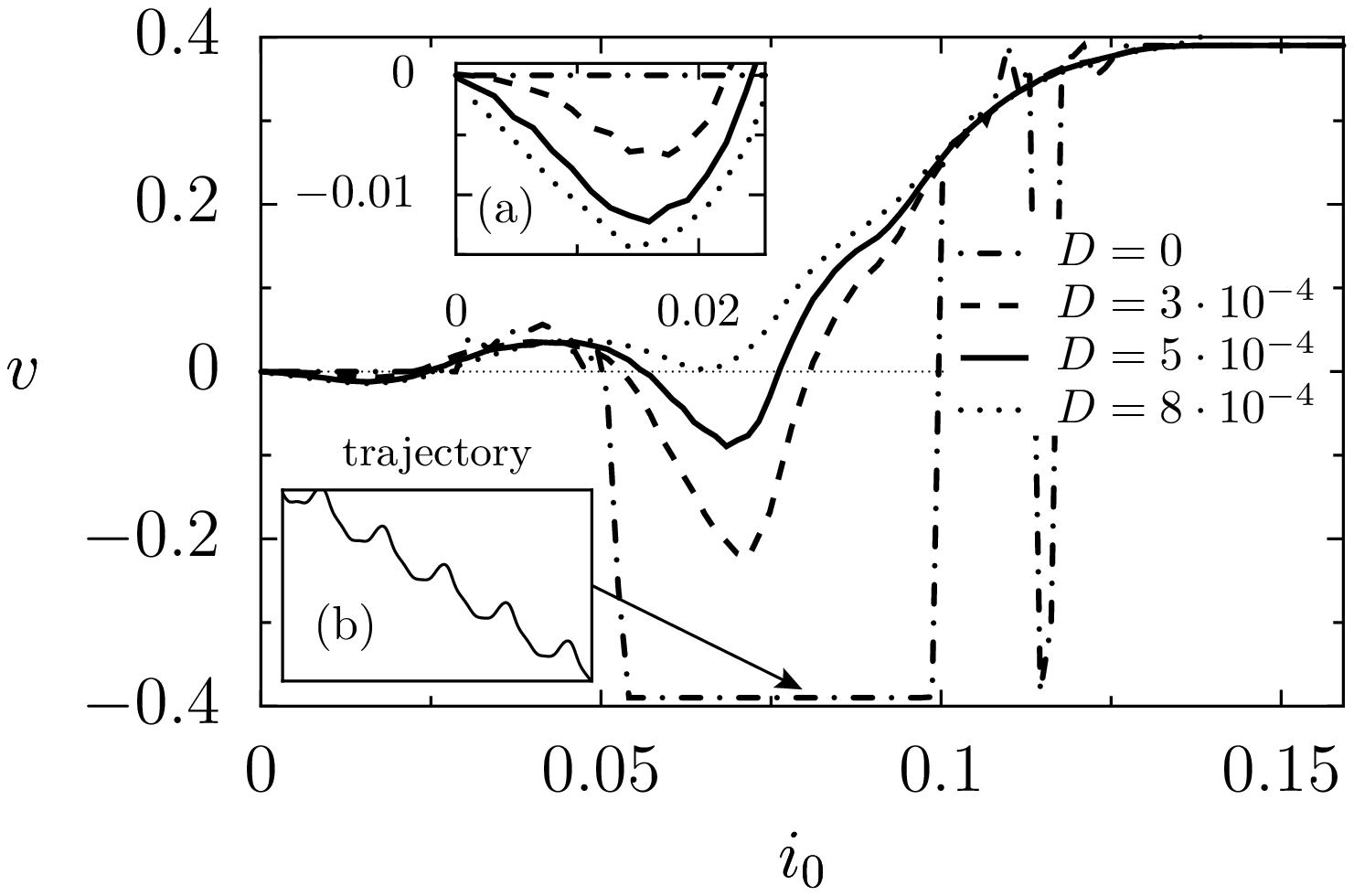}
\caption{Voltage  {\it vs} dc-current:  The re-entrant effect of the
negative response ($v<0$) is shown at a positive bias ($i_0>0$) for
the  following set of  parameters:  $i_1=0.668$, $\sigma=0.143$,
$\Omega_1=0.78$ and four values of temperature $D$. For small $i_0$,
the absolute negative conductance is observed for non-zero
temperature and it  is solely induced by thermal fluctuations, cf.
panel (a).   For larger $i_0$, negative  nonlinear conductance
results as a deterministic effect. The corresponding  long-time
trajectory  $\phi(t')$  for $i_0=0.0796$ shown in panel (b)  belongs
to running states into the negative direction of $\phi$  resulting
in the large  average voltage. The value of the amplitude of the
ac-driving corresponds to the  first negative-valued  minimum of the
mean voltage with respect to the ac-current, cf. in Fig. \ref{adm}.
} \label{reentrant}
\end{figure}
In  the linear response  regime, the voltage tends to zero when the
dc-current tends to zero. In the nonlinear response regime, the
voltage can tend to zero even if the dc-current assumes a non-zero,
finite value. For example, there can emerge two  positive values of
the dc-current $\{i_{01}, i_{02}\}$ such that in-between the voltage
is negative. Then, the nonlinear conductance coefficient $g_N(i_0)$
is negative-valued  in this very interval $(i_{01}, i_{02})$;
meaning that   the voltage takes on the opposite sign of the
dc-current for a non-zero valued dc-bias strength $i_0$. Indeed, this
situation is presented with Fig. \ref{reentrant}. The amplitude of
the ac-current is set at $i_1=0.668$; it corresponds to the first
negative-valued minimum of the averaged voltage depicted in Fig.
\ref{adm}. For small dc-currents, ANC exists (see inset of Fig.
\ref{reentrant}) while for larger dc-currents $i_0$, negative-valued
nonlinear conductance (NNC) results. NNC is a predominantly
deterministic phenomenon which survives in presence of small thermal
noise. Indeed, we observe in Fig. \ref{reentrant} that for the case
$D=0$, this effect is most pronounced and takes place in a wide
parameter interval $(i_{01}, i_{02})$. The long-time trajectories of
the dynamical system (\ref{JJ2}) correspond to running states into
the negative direction of $\phi$, which are of period two and
therefore the deterministic average voltage is large, cf. Fig.
\ref{reentrant}.  When the temperature increases, the interval
$(i_{01}, i_{02})$ shrinks and the amplitude of the voltage
decreases. Above some temperature ($D \approx 8\cdot 10^{-4}$ in
Fig. \ref{reentrant}), the NNC  effect disappears and only ANC
survives.

We end this section by a statement that the occurrence of anomalous
transport  may be governed by different mechanisms. In some regimes
it  is  solely induced by thermal equilibrium fluctuations, i.e. the
effects are absent  for vanishing  thermal fluctuations $D=0$. In
other regimes, anomalous transport may also occur in the noiseless,
deterministic system and the effects increasingly fade out  with
increasing temperature \cite{machura}. Both situations are rooted in
the complex deterministic structure of the nonlinear dynamics
governed by a variety stable and unstable orbits.

\section{ Optimal parameter regimes for the occurrence of negative conductance}
\begin{figure}[htb]
  \begin{center}
    \includegraphics[angle=0,width=0.95\linewidth]{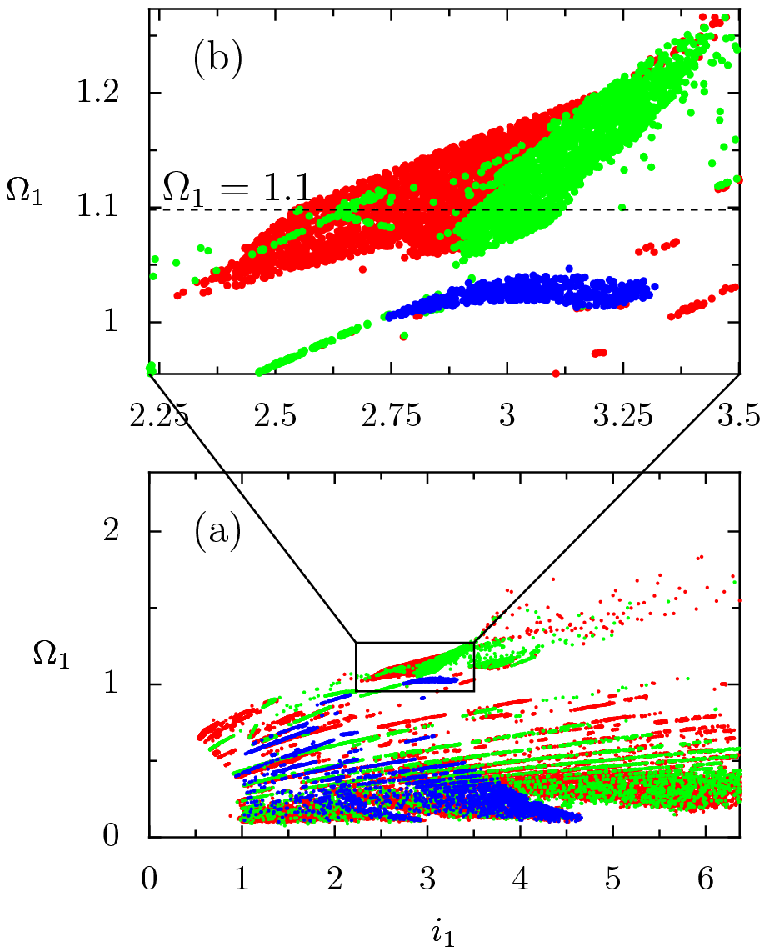}
  \end{center}
  \caption{(color online)   The transport properties of the driven system in the parameter space
    $\{i_1, \Omega_1\}$  at a representative friction value of
    $\sigma=0.191$,  dc-bias $i_0=0.0159$
   and zero noise strength $ D=0$.  All points in the parameter space, where  a
   negative conductance occurs,
   are marked by symbols with different grey scale. The coding
   corresponds to different regimes of values assumed by the
    ratio  $v/\Omega_1$:  dark grey (online: red) denotes the interval
    $(-0.94, 0)$, light grey (online: green): $(-1.26, -0.94)$, and black
    (online: blue) corresponds to values less
    than $-1.26$.
Most of  the dark grey points correspond to chaotic trajectories.
 It     turns out that the  black regimes are most susceptible to thermal
fluctuations, meaning that the negative-valued conductance rapidly
fades away towards positive values with increasing  thermal noise
intensity. In panel (a)     one clearly can distinguish a large
compact region exhibiting negative-valued
     conductance, being enlarged with the upper panel (b).  In
     regions different from this one we find that negative-valued
     conductance occurs in narrow  "bands'' only.
      For an experimental realization of negative conductance the most
    promising parameter regimes are those  regions marked by  light grey. This
    holds in
    particular for the zoomed light grey regimes depicted in panel (b).}
  \label{fig4}
\end{figure}

In our recent work in Ref.  \onlinecite{machura} and in the previous
section, we discussed some fixed parameters  for $\{\sigma,
\Omega_1\}$ for which the negative conductance  (ANC or/and NNC)
does emerge. Are these two values exceptional? To answer this
question, we have searched the part of the three dimensional
parameter space that is specified by ac-driving strength $\{i_1 \in
(0,6.37)\}$, angular driving frequency $\{\Omega_1 \in (0,1.91)\}$,
friction strength $\{\sigma \in (0,0.796)\}$ in order to locate
regions of negative conductances  in the deterministic case, when
$D=0$. We set the dc-current to a small value $i_0=0.0159$ and
calculated the value of the current for many randomly chosen
parameters $i_1$, $\Omega_1$ and $\sigma$. \cite{footnote} The
inspection of the results revealed that there are values of friction
$\sigma$  for which regions of negative conductance are most
prominent. One such value is $\sigma=0.191$ for which the negative
conductance is most pronounced in a relatively large domain of
parameter variation with relatively large values of the
dimensionless voltage. Therefore we performed a more accurate search
for the section $\sigma =0.191$ and analyzed the two-dimensional
parameter domain $\{i_1 \in (0,6.37) , \Omega_1 \in (0,1.91)\}$. The
results  are depicted with Fig. \ref{fig4}.
 The points in Fig. \ref{fig4} which are indicated by different grey scales
correspond to values where $v(i_0=0.0159)<0$: the voltage assumes
thus the opposite sign  of the dc-current.  Strictly speaking a
negative voltage at the finite bias $i_0=0.0159$ does not
necessarily imply ANC. However, in most of cases, under these
conditions ANC is observed  in the presence of small thermal noise.
We have also found, that the noise induced ANC  typically occurs
also for parameters which differ only slightly from those where
$v(i_0=0.0159)<0$. Thus, this systematic analysis  provides insight
into the structure of the parameter space and proves useful  for
designing corresponding Josephson junction experiments. Finally,
using extensively this technique we are rather convinced that no
remaining regimes of the negative conductance are likely to emerge
in this system than those already depicted in Fig. \ref{fig4}.
 The grey scales  in Fig. \ref{fig4}
represents various  regimes of the ratio $v/\Omega_1$. One can
distinguish two features in the parameter space:  a stripe-like
structure for a broad range of parameters and one pronounced region,
 being zoomed in  panel (b). The stripe-like structure suggests that in an
experiment,  one will observe with large probability negative
conductance upon variation of a single parameter.  In this context the
region shown in panel (b) is especially
interesting  because the negative conductance depicted there
exhibits a relatively  large robustness with respect to a variation
of the system parameters.

\begin{figure}[htb]
  \begin{center}
    \includegraphics[angle=0,width=0.95\linewidth]{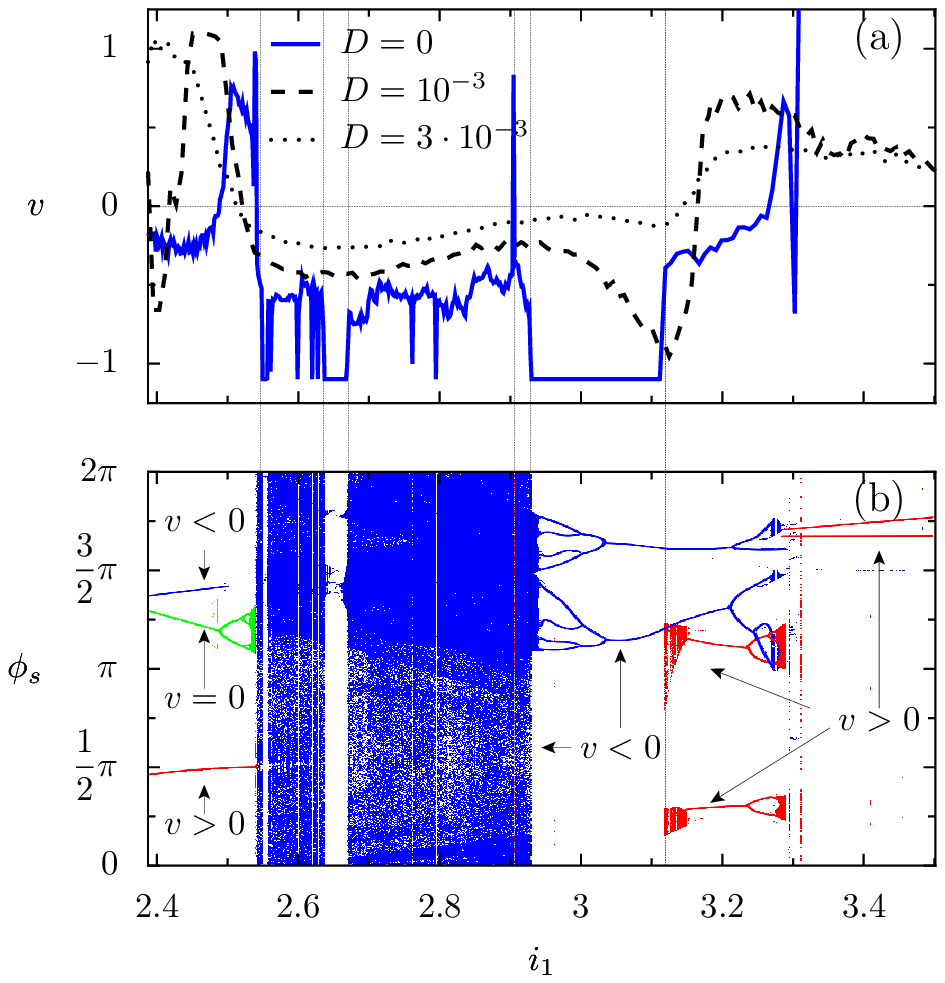}
  \end{center}
  \caption{( color online)
This figure  elucidates  the  transport features and the
    dynamical properties for parameter values in  $\{i_1,\Omega_1\}$-
    space within the section obtained by fixing the angular
    driving frequency at $\Omega_1=1.1$, cf. the dotted line in panel (b) of  Fig. \ref{fig4}.
    The blue (solid)
    line in panel (a)  depicts the dependence of  the average voltage on
    the amplitude of the ac-current for zero noise $D=0$.
    The influence of thermal fluctuations on the average voltage is shown for
     two temperature values $D=10^{-3}$  (dashed line) and
    $D=3\cdot 10^{-3}$ (dotted line).   We do find  that   a negative-valued conductance
     indeed can survive at small non-zero temperatures.
    The transport properties of the system follow from  the
    underlying complex dynamics. In panel (b) we show the bifurcation diagram:
    the Poincare section of the phase in the deterministic system.
    The attractors are color coded according  to the sign
    of the assumed corresponding average voltage: blue yields a
    negative average voltage (also pointed by arrows with $v<0$), green  gives zero
(also pointed by arrows with $v=0$) and red
    amounts to a positive average voltage (also pointed by arrows with $v>0$).
}
  \label{fig5}
\end{figure}
In  the deterministic case, we identify three classes of trajectories
which generate  negative conductance.
 This  is  marked by  different grey scales in Fig. \ref{fig4}.
 Dark grey corresponds to chaotic trajectories.  For this regime of
 parameters the conductance is negative also for small non-zero temperatures.
 Light grey corresponds to periodic orbits. For this regime of parameters
 the conductance is  negative   also for small non-zero temperatures.
Black regions also correspond to periodic orbits; for this regime of
parameters the negative conductance however quickly diminishes as
the temperature is raised.

From the experimental view-point,  the most interesting regimes are
those where the negative conductance is most  robust within some
finite temperature interval.   In the case presented in Fig.
\ref{fig4}, this corresponds to  the dark and light grey regions.
Therefore, we set e.g.  $\Omega_1 =1.1$, cf.  panel (b) of Fig.
\ref{fig4},   in order to observe the variation of  the negative
conductance when passing through the dark  and light grey regions.
Details are depicted in the bifurcation diagram, see Fig.
\ref{fig5}. In the deterministic case, the voltage remains almost
constant under  small variations of the amplitude $i_1$ of the
ac-current  when  the parameters  belong to the light grey regions,
note the voltage dependence on $i_1$ around $i_1=2.64$ and $i_1=3$
depicted in  panel (a)  of  Fig. \ref{fig5} for $D=0$. On the other
hand, the voltage changes irregularly when $i_1$ is changed smoothly
within the red regions, note the voltage dependence on $i_1$  around
$i_1=2.78$ depicted in panel (a)  of  Fig. \ref{fig5} for $D=0$.  In
Fig. \ref{fig5}, the averaged voltage is also shown for a small but
non-vanishing temperature. When driving within the interval $a\in
(2.55, 3.15)$, the voltage is negative  for $D=10^{-3}$ and
relatively stable with respect to a small variation of parameters.
The region around $i_1=3$ lends itself as optimal  for an
experimental verification of our predicted anomalous transport
features.


\section{ Experimentally accessible regimes for  anomalous transport}
The results presented above are given for dimensionless variables
and dimensionless parameter values. In order to motivate
experimentalists to test our predictions and findings, it is
convenient to transform all quantities back to their original
dimensional values and dimensional parameter strengths. There are
three important parameters which characterize the function of a
Josephson junction. These are the critical current $I_0$, the
resistance $R$ and its capacitance $C$. Three further parameters
characterize the external driving, namely the strength of the
dc-current $I_d$, the amplitude of the periodically varying
ac-current $I_a$ and its angular frequency $\Omega$.
Finally, the temperature $T$ must be
chosen large enough such that the junction operates in the
semi-classical regime. These  physical quantities are related to the
corresponding  dimensionless quantities by the relations:
\begin{eqnarray} \label{tran1}
 V= \left(\frac{\hbar \omega_p}{2e} \right) v, \quad
\Omega = \omega_p  \Omega_1, \quad \frac{1}{RC} = \omega_p \sigma
\;,
\end{eqnarray}
with the frequency scale $\omega_p$ given by the characteristic
plasma frequency, reading
\begin{eqnarray} \label{omega}
\omega_p^2 = \frac{2e I_0}{\hbar C}\;.
\end{eqnarray}
The driving strengths and the actual physical temperature read
\begin{eqnarray} \label{tran2}
I_a =I_0  i_1, \quad
I_d=I_0  i_0, \quad T = \left(\frac{\hbar
I_0}{2 e k_B} \right) D \;,
\end{eqnarray}
being all scaled  by the value of the critical current $I_0$.

First, one  should fix the operational temperature $T$ of the
classical experimental regime.   The   last relation in
(\ref{tran2}) then yields the strength of critical current $I_0$.
This in turn determines the  amplitude strengths  $I_a$ and $I_d$.  Fixing
 the frequency $\nu=\Omega/2\pi$ of the microwave source, then
determines via the  chosen relevant value of the dimensionless
parameter $\Omega_1$ the strength of the plasma frequency $\omega_p$.
This in turn determines the magnitude of the capacitance $C$ via the
relation in (\ref{omega}) and the resistance $R$  follows from the
last relation in (\ref{tran1}). Moreover, one should check
whether the parameters such chosen  obey the
inequalities:
\begin{equation}
\label{clas}
\hbar \omega_p << E_J, \quad \hbar \omega_p << k_B T,
\end{equation}
which taken together guarantee that the junction indeed operates in
the  (semi)-classical regime as presumed with the model equation  in
(\ref{JJ1}). These inequalities imply that the level spacing of the
plasma oscillation is small, both compared to the coupling energy
$E_J$  and the thermal  energy $k_B T$.

To be explicit, we here evaluate some real experimental
circumstances which can  present  "optimal" conditions to
experimentally verify  negative conductance. Upon inspection of our
dimensionless analysis the following parameter set is an example of the
optimal regime: $i_1=3, \,\Omega_1 =1.11, \,\sigma =0.191$ and
$D=10^{-3}$. For a physical temperature of $T=4$K, the critical
current  then is $I_0=167.8 \mu$A, the amplitude strength  becomes
$I_a=507 \mu$A, the ac-angular frequency  emerges as $\Omega =
150$GHz,  $I_d=2,67 \mu$A with the capacitance being $C=27.9$ pF and
the resistance value at $R =1.4 \Omega$. Under these conditions, the
absolute value of the voltage amounts to $V=3.54 \mu$V.

The NDC regime, which is presented in Fig. \ref{ndm},  is observed
for  $i_1= 0.73, \,\Omega_1 =0.78, \,\sigma = 0.143$ and $D=10^{-3}$
which for a temperature at $T=4$K implies a  critical current
$I_0=167.8 \mu$A, $I_a=122 \mu$A,  ac-angular frequency $\Omega =
150$GHz, and $I_d$ varying between  $\in (13.36, 14.43) \mu$A for
$i_0 \in (0.0796,0.086)$, with the capacitance set at $C=13.83$ pF
and the resistance set at $R = 2.63 \Omega$. Under these conditions,
the absolute value of the voltage  approximately reads $V=730 \mu$V.

\section{Conclusions}

With this work we took a closer look at the richness of anomalous
transport behavior occurring in a biased and harmonically driven
Josephson junction. As it turns out, the  underlying chaotic
dynamics together with the  influence of thermal noise
triggers a whole new variety of unexpected transport features. Apart
from regions displaying negative differential conductance behavior
we could identify novel transport characteristics such as
noise-induced absolute negative conductance near zero bias and
negative-valued conductance in the strongly nonlinear response
regime.   Let us summarize these  various transport phenomena occurring
in an ac-driven  Josephson junction as described by Eq. (\ref{JJ1}):
\setdefaultenum{(1)}{}{}{}
\begin{compactenum}
\item The  dependence of the voltage on  the angular driving frequency
  $\Omega_1$  of the ac-current drive
depicts  windows of deterministic and thermally induced ANC regimes:
ANC appears and disappears as the frequency increases (for a
preliminary account on this effect  see also Ref.  \onlinecite{tokio}).
\item The  dependence of voltage versus the amplitude $i_1$ of the
ac-current drive depicts windows of thermal-noise induced ANC
regimes, cf. Fig. \ref{adm}.
\item The  voltage behavior at fixed bias as a function of thermal
  temperatures  $D$ exhibits many
familiar features known from the field of Brownian motors
\cite{motor},  such as the occurrence of a voltage
reversal versus D, or a typical   bell-shaped behavior versus noise
strength D.   \cite{ratchet}
\item The voltage as a function of  the bias current $i_0$  can  exhibit each of
the three anomalous transport features, namely
 absolute negative conductance (ANC),
negative-valued nonlinear conductance  (NNC)  and  negative
differential conductance (NDC).
\item Reentrant phenomena  of negative conductance regimes occur as a function of the dc-bias $i_0$:
Starting out from zero, the voltage may decrease  for increasing
$i_0$, reaching a negative-valued local minimum  value which changes
over into a local positive-valued  maximum upon increasing $i_0$.
Upon further increasing $i_0$, the voltage   starts to decrease again into a local,
negative-valued minimum, thus exhibiting NNC. Finally, it
increases monotonically with increasing $i_0$,  displaying an almost
perfect Ohmic-like dependence.

\end{compactenum}

Our identified novel transport features as presented in Fig.
\ref{adm} and Fig. \ref{fig5} are also accessible to an experimental
verification via appropriately designing the experimental working
parameters for the Josephson system. Here we identified such
parameter sets in our section V while  yet a different one has been
indicated in our earlier presentation in Ref. \onlinecite{machura}.
We are confident that our predictions will invigorate
experimentalists to undertake the experimental efforts to check our
various predictions.

\begin{acknowledgments} Work supported by the DFG via grant HA
1517/13-4, the DFG-SFB 486,  the German Excellence Initiative via
the \textit {Nanosystems Initiative Munich (NIM)},   the grant
MNiSW  N 202 131 32/3786 and the DAAD-MNiSW program "Dissipative
transport and ordering in complex systems".
\end{acknowledgments}

\section*{Appendix}

In our previous work \cite{machura} we used a scaling popular among
the 'ratchets' community, where the 'coordinate' variable
$x=\phi/2\pi$ is rescaled to the unit interval.  Then  the rescaled
form of Eq. (\ref{JJ1}) reads
\begin{equation}
\label{JJ3}
\ddot{x} + {\gamma} \dot{ x} +2\pi  \sin (2\pi x)  = f + a \cos(\omega s)
+ \sqrt{2 \gamma D} \; \Gamma(s),
\end{equation}
where the dot denotes differentiation with respect to the
dimensionless time $s=t/\tau_1$,  where $\tau_1 = 2\pi/\omega_p$.
 The relations between the  parameters in two scalings in eq. (\ref{JJ2}) and (\ref{JJ3})  are as follows:
\begin{equation}
\label{par} \gamma = 2\pi \sigma, \quad f=2\pi  i_0, \quad a=2\pi
i_1, \quad \omega = 2\pi  \Omega_1
\end{equation}
Now, the  dimensionless velocity $v = \langle dx/ds \rangle$  and
expressions  for the  voltage in Eq. (4) and  the noise intensity  $D$ are identical
for  both scaling procedures.


\begin{thebibliography}{99}
\bibitem{footnote} We have used
 the  following percolation-like algorithm: starting from a point where
  the voltage  is negative, we are perturbing the parameters by small
  random values and repeatedly calculate the  voltage. When the local
  density of points with a negative voltage exceeds some prescribed
  value,  a larger random perturbation is applied, until  we  find a
  new region
  with a  negative  voltage. Then again a search with small random
  perturbations
  was performed.  Such an algorithm can be easily implemented
  on a PC-cluster,
  and yields a much better exploration  of the  parameter space than
  by a deterministic search on a
  regular mesh, especially in higher dimensions and
  for unknown ranges of parameters.
\end{thebibliography}
\end{document}